\documentclass[mathleft
]{an}
\usepackage{graphicx}
\usepackage{times}

\begin{document}
\hyphenation{}
\Pagespan{590}{603}
\Yearpublication{2015}%
\Yearsubmission{2015}%
\Month{Aug}%
\Volume{336}%
\Issue{6}%
\DOI{10.1002/asna.201512195}%
\def\arcsec{\hbox{$^{\hbox{\rlap{\hbox{\lower4pt\hbox{$\,\prime\prime$}}
          }\hbox{$\frown$}}}$}}

\title{The Bochum Survey of the Southern Galactic Disk: \\
        II. Follow-up measurements and multi-filter photometry for 1323 square degrees monitored in 2010\,--\,2015}

\author{Moritz Hackstein\inst{1}\fnmsep\thanks{Corresponding author:
  \email{hackstein@astro.rub.de}\newline}
\and Christofer Fein\inst{1}
\and Martin Haas\inst{1}
\and Michael Ramolla\inst{1}
\and Francisco Pozo Nu\~{n}ez\inst{1}
\and Angie Barr Dom\'inguez\inst{2}
\and Lena Kaderhandt\inst{1}
\and Karl Thomsch\inst{1}
\and Nina Niedworok\inst{1}
\and Christian Westhues\inst{1}
\and Rolf Chini\inst{1,2}
}
\titlerunning{Southern Galactic Disk Survey II}
\authorrunning{M. Hackstein et al.}
\institute{
Astronomisches Institut,
Ruhr--Universit\"at Bochum,
Universit\"atsstra{\ss}e 150,
D-44801 Bochum, Germany
\and
Instituto de Astronom\'{i}a,
Universidad Cat\'{o}lica del
Norte, Avenida Angamos 0610, Casilla
1280 Antofagasta, Chile
}

\received{2015 Mar 13}
\accepted{2015 Jun 08}
\publonline{2015 Aug 01}


\abstract{%
  This paper is the second in a series describing the southern Galactic Disk Survey (GDS) performed at the Universit\"atssternwarte Bochum near Cerro Armazones in Chile. Haas et al. (2012, Paper~I) presented the survey design and the characteristics of the observations and data. They identified $\sim\!\!2200$ variable stars in an area of 50 square degrees with more than 50 observations in 2011.\\
  Here we present the first complete version of the GDS covering all 268 fields with 1323 square degrees along the Galactic disk including revised data from Paper~I. The individual fields were observed up to 272 times and comprise a maximum time span between September 2010 and May 2015. We detect a total of 64\,151 variable sources, which are presented in a catalog including some of their properties and their light curves. A comparison with the International Variable Star Index (VSX) and All Sky Automated Survey (ASAS) indicates that 56\,794 of these sources are previously unknown variables. Furthermore, we present $U$, $B$, $V$, $r'$, $i'$, $z'$ photometry for all sources within the GDS, resulting in a new multi-color catalog of nearly $16\!\times\!10^6$ sources detected in at least one filter.\\
Both the GDS and the near-infrared VISTA Variables in the Via Lactea survey (VVV) complement each other in the overlap area of about 300 square degrees enabling future comparison studies.}

\maketitle

\section{Introduction}

During the last decade a number of projects have emerged that address the astrophysical time domain by wide-field multi-epoch surveys with dedicated telescopes and the corresponding equipment. Prominent examples are the All Sky Automated Survey (ASAS, Pojmanski 2002; Pojmanski \& Maciejewski 2004), the Optical Gravitational Lensing Experiment (OGLE, Udalski et al. 1992), both founded by Bohdan Paczy\'nski, the MACHO collaboration (Alcock et al. 2000), the Catalina survey (Drake et al. 2009) and the space based missions CoRoT and Kepler. These are accompanied by remarkable amateur monitoring cataloged in the general and extensive data base of the AAVSO International Variable Star Index (VSX\footnote{http://www.aavso.org/vsx/}, Watson et al. 2011).\par
One of the current surveys is the VST Photometric H$\alpha$ Survey of the Southern Galactic Plane and Bulge (VPHAS+) within the latitude range $-5^\circ < b < +5^\circ$. It provides data at $u, g, r, i$, and H$\alpha$ down to $>$\,$20^{th}$ magnitude with a seeing-limited resolution of about 1''. However, the saturation limit is around $12^{th}$ magnitude, thus missing the apparently bright portion of the Galactic population. The situation becomes even worse with the upcoming LSST project where sources as faint as $\sim 16^{th}$ magnitude start saturating. On the bright side there is the ASAS project which has produced extensive catalogs of variable stars (ACVS) of the southern hemisphere (Dec $< +28^\circ$). The unsaturated brightness range is roughly $\sim 7 < V,I < 13$\,mag, however, with a low spatial resolution of only 14.3''/pixel. Therefore, a complementary survey is desirable that extends the VST and LSST surveys of the southern Galactic plane toward brighter stars at a higher angular resolution of typically 3'' in case of the Bochum Galactic Disk Survey (GDS). For a detailed discussion of resolution and source separability see Paper~I.\par
The GDS is an on-going project dedicated to monitor the variability of the intermediately bright ($7 < r', i' < 18$) stellar population in a 6$\degr$ wide stripe along the Galactic plane. During the same year span as the GDS (since about 2009), the near-infrared (0.9--2.5$\mu$m) Vista Variables in the Via Lactea survey (VVV, Minniti et al. 2010)  monitors 520 square degrees of the southern Milky Way, compared to 1323 square degrees surveyed by the GDS. The GDS stripe is about 220$^{\circ}$\,$\times$\,6$^{\circ}$ long/wide, the VVV stripe 55$^{\circ}$\,$\times$\,4$^{\circ}$ long/wide plus a 20$^{\circ}$\,$\times$\,15$^{\circ}$ long/wide box at the bulge. The $K$ band brightness ranges from 9.5\,mag (saturation limit) to 18\,mag, hence nicely covers the $r'$ and $i'$ brightness range of the GDS for ``blue'' sources ($r'-K \approx 0$). However, red sources in the GDS have $r'-K \approx 5$ ($K_s$ from 2MASS) and $r'<$ 15.5\,mag, so that unfortunately many of them may be saturated in VVV's $K$ band images. Nevertheless, also despite the different spatial resolution, the surveys GDS and VVV are (partly) complementary, and well worth future comparison studies in the overlap area of about 300 square degrees (75$^{\circ}$\,$\times$\,4$^{\circ}$).

Meanwhile all 268 fields (designated as ``GDS hhmm-ddmm'' of central right ascension and declination, respectively) of the mosaic have been observed, 220 of them with 20 to 272 measurements in $i'$ and $r'$. Compared to Paper~I, we extended the survey by observing all fields additionally at least once in $U$, $B$, $V$ and $z'$. These new observations are still on-going and will be completed within 2016; the current state of this multi-color approach is included in this paper. By extending the filter set we yield multiple measurements across the visible spectrum allowing a determination of classical colors like $U\!-\!B$, $B\!-\!V$, $r'\!-\!i'$ and $i'\!-\!z'$ and to obtain a rough spectral energy distribution. In this way, the temperature of interesting sources as well as other properties, like visual extinction or excess toward longer wavelengths can be estimated.

This second release of the GDS has been obtained by our refined pipeline -- with no necessity of manual inspection of the results -- while yielding a very low rate of false positives. Furthermore, our analysis methods have been extended and refined compared to the first release in Paper~I.

The GDS project aims at detecting any type of variable objects, that is, objects changing flux over time. Therefore, we strive to refine our identification method to be biased as little as possible toward the type of variability. As described in Paper~I, we use three simple detection methods, which together do not presuppose periodicity or other decisive properties characteristic of any particular type of variable star.

\section{Observations}

Compared to Paper~I we extended our survey during the years 2011 to 2015, now covering all 268 fields (1323 square degrees) with at least one combined image. The fields are positioned on a 2$^{\circ}$ grid. Other than that, we continued the same observing strategy, i.e., performing simultaneous 10\,s exposures cropped to $2.2^{\circ}\!\times2.2^{\circ}$ in the Sloan $r'$ and $i'$ band filters. Of those exposures a series of nine dithered images is taken subsequently of the same field and later combined during data reduction. The GDS comprises stars in the range $8^{m}\!< \!r' \!< \!18^{m}$ and $7^{m}\! < \!i' \!< \!17^{m}$.

\section{Data reduction}

The data reduction was described extensively in Paper~I and has not changed significantly. However, by re-analyzing the increased data set we found that the SCAMP contrast parameter provided by the SCAMP tool from the AstrOmatic\footnote{https://www.astromatic.net/} soft\-ware suite (Bertin \& Arnouts 1996; Bertin 2006) is helpful in assessing the image quality of our raw data in advance to further analysis. This parameter ``is defined as the ratio of the amplitude of the detected peak (within the allowed limits in pixel scale and position angle) to the amplitude of the second highest peak found in the cross-correlation. In practice, contrast factors above $\approx$\,2 shall be considered as reliable, while contrasts below 2.0 are dubious'' (from the SCAMP manual\footnote{\label{astromatic}https://www.astromatic.net/pubsvn/software/scamp/trunk/doc/scamp.pdf}). We found that a contrast parameter $<$\,2 in our data usually indicates pointing errors or other severe problems in our raw data. This data is then discarded. Additionally, we discovered that at certain telescope orientations the shutter from the camera at RoBoTT-A did not close completely. This issue has been fixed and influenced only data sets before April 2014, which have been inspected and selectively removed.

\section{Construction of light curves}

We have extended the pho\-to\-me\-try and analysis process based on our experience with the first data set published in Paper~I. Since almost every night new images are collected, the GDS is designed to be self-updating and hence to yield a rolling release of all available data. To accomplish this task, a fast pipeline work-flow is needed which produces results that do not have to be checked individually after each update process.

For each field and filter the data consist of typically one combined image per night. For an increasing number of fields we also added multiple observations per night to increase our sensitivity toward short-term variability. The light curve construction of all sources is based on the procedure described in Paper~I with the following changes and additions:

\begin{itemize}
\item [--]
Check number of single images combined to one reduced image. Usually an image is the combination of nine single dithered exposures.  We chose a minimum of seven raw exposures for one image to be analyzed further.

\item [--]
Check SCAMP contrast parameters of the single exposures. Manual inspection showed that images with contrast values $<$\,2 are severely flawed and not usable. Therefore, we only consider images of which less than two of the single frames show low contrast values before combination for further analysis.

\item [--]
Source detection on each image is accomplished using SourceExtractor (SE) from the AstrOmatic software suite in single image mode. By that we gain parameters for quality control uninfluenced by the master image that is later used in double image mode.
We derive flagging parameters for each night based on the total number of detected sources (declining in case of poor transmission, strong wind, and bad focus), average FWHM, average source elongation,
and the moon distance.
Images that show less than 70\% of the median number of stars found in all images of this particular field and filter are discarded.
Those images with very poor stellar profiles (i.e., elongation or FWHM exceeding 3\,$\sigma$ of all nights) are omitted as well.

\item [--]
Average combination of the 12 best images that passed quality control to a common master image (average with rejection of the lowest and highest pixel value). This number has proven to be a good compromise between master image quality and computation time. The more images we combine the lower is the noise of the resulting master image. Combining a master image provides better average center coordinates for all sources and improves the separation of close faint sources considerably. The better center positions improve the photometry of faint sources only slightly, since the photometry is still done on single frames, which keep their original noise levels. Hence, combining a significantly larger number of images for the master image is not necessary and we set a high detection threshold of 10\,$\sigma$ in the master image to account for the lower $S/N$ ratio in the single images. If less images are available, all images are used. By that we obtain a
master image with low noise, good averaged source positions  and well sampled point spread functions. The latter minimizes the influence of elongated sources and bad seeing in single images during the following double image mode photometry.

\item [--]
Source detection on each image using SE in double image mode. By using the double image mode we avoid the problem of cross-matching extremely large data sets, which would require unrealistically high amounts of CPU time. On the other hand, our tests have shown that due to varying image quality it happens ever so often that either close sources are mismatched in more crowded areas or sources that are separated in good images are merged and analyzed as a single source in images with poorer seeing. We deal with those issues by using the double image mode with ''AUTO'' photometry that uses a fixed aperture center position and Kron radius for each star over all nights. The Kron radius yields elliptical apertures determined automatically based on the shape, elongation and orientation of the individual source  (Kron 1980). In case of crowded regions where light from a neighboring source is detected within the aperture, it is removed by interpolating with the opposing side of the aperture. If this is not possible, the source is removed. This does not account for varying seeing, but still has proven to produce more stable results than other methods, including PSF photometry, for our data. We then dismiss all sources that are either saturated in the master image or lie below the master image's completeness limit. We also store quality flags provided by source extractor (SE flags),  which specify the quality of a photometric measurement with regard to separability of close sources, bad pixels, and saturation. The airmass correction follows Patat et al. (2011) for Cerro Paranal.

\item [--]
The overall magnitude calibration is done as described in Paper~I using Landolt standard stars (Landolt 2009).
Usually, the absolute flux calibration of a target star field
is achieved by observing photometric standard star fields before and after the target. However, this is extremely time consuming. Therefore we performed a different ``statistical'' strategy. Per night only three photometric Landolt standard star fields are observed, at the beginning, middle and end of a night.  In a first step (see below) for each GDS and Landolt star field the ``relative'' light curves of the stars are created in instrumental magnitudes. These  light curves are calculated after correcting each observation to airmass $z$\,=\,1 and correcting for transmission differences between different nights. Thus the resulting relative light curves in instrumental magnitudes belong to optimal photometric conditions. This procedure is done in the same manner for both the GDS fields and the Landolt fields, both observed with same filters and exposure times.\par
The filters of the GDS are Johnson \textit{UBV} and Sloan $r'i'z'$, their properties are listed in Table \ref{tab:filters}. The Landolt catalogs list \textit{UBVRI} (Johnson-Cousins) photometry, henceforth denoted as catalog magnitudes. For the \textit{UBV} filters we use these catalog magnitudes to derive the offset to the instrumental magnitudes, and thus the absolute photometry of stars in the GDS fields.\par
For the Sloan $r'i'z'$ filters the same procedure is applied using a transformation of  Landolt catalogs to Sloan $r'i'z'$. To this point, we did not use (color-dependent) conversion formulas. Instead, we converted for each Landolt star its \textit{BVRI} photometry into spectral energy distribution (SED, e.g., in units of  mJy, see Table \ref{tab:filters}) and fitted the SED with a Planck function. For each filter this fit is excellent for $>$\,80\% of the Landolt stars. We then only use those Landolt stars where the difference between catalog SED and Planck fit is less than 1\% in all four filters (\textit{BVRI}). The smooth Johnson-Cousins broad band SED strong\-ly suggests that the Landolt star has no strong spectral feature in the Sloan broad bands $r'i'z'$. Then, the Sloan $r'i'z'$ SED values are taken from the Planck fit and further converted to magnitude, using the values in Table \ref{tab:filters}. In this fitting procedure the \textit{U} filter is not used because many stars show spectral features shorthand of 4000\,\r{A}.\par
The main purpose of the GDS is to identify variable stars. Many of them have large amplitudes up to several magnitudes. In addition, some (if not many) stars are observed mainly at low or high brightness. Thus, for us the precision of the absolute photometry plays a minor role. In any case the relative light curves (in instrumental magnitudes) already allow us to recognize, for instance, whether a star  gets bluer when brighter etc. Therefore, we consider the precision of the flux calibration as sufficient, and we give a rather conservative large global absolute photometric uncertainty of up to 0.1 to 0.2\,mag to avoid over-interpretation of the photometric accuracy in further studies.\par
We find that our ``statistical'' photometric calibration procedure works well if a field has been observed often, in our experience in more than about 10 nights in that particular filter. For instance, the SEDs of the millions of non-varying stars show a strikingly smooth shape, which in most cases ($>$\,90\%) is excellently fit by a Planck function, similar as for the Landolt stars. The detailed analysis of the procedure, including the effect of color-dependence, will be presented in a future work.

\begin{table}
	\centering
		\begin{tabular}{rcccc}
			\hline
Filter & Central wavelength ($\mu \mathrm{m}$) & $F_0$ (Jy) \\
\hline
$U$ &   0.365  &      1895.8\\
$B$  &  0.433   &     4266.7\\
$V$ &   0.550   &     3836.3\\
$r'$ &   0.623  &      3631.0\\
$i'$ &   0.764  &      3631.0\\
$z'$ &   0.906  &      3564.7\\

			\hline
		\end{tabular}
	\caption{Filter properties used to derive the transformation between
Johnson-Cousins and Sloan magnitudes taken from {http://www.gemini.edu/?q=node/11119}}
	\label{tab:filters}
\end{table}

\item [--]
We use all stars with good pho\-to\-me\-tric quality (i.e., all zero-flagged and therefore unsaturated isolated sources above the completeness limit) as relative calibrators for transmission and image quality to account for night-to-night fluc\-tu\-a\-tions as described in Paper~I. We find that usually the brightness dependency of this calibration factor is nearly linear in magnitudes for most nights. Hence, we use a linear least squares fit for each half-magnitude bin to determine the calibration offset to flatten the calibrator light curves on average. The result is shown in Table~\ref{tab:calib}. The brightness dependency of the night-to-night offset seems to be mainly due to the brightness dependency of the FWHM and the overestimated background brightness due to unresolved sources. The former is a result of non-optimal focusing as well as changing weather conditions. The source elongation originates from higher wind speeds and tracking inconsistencies for certain telescope orientations. Since the double
image
mode photometry uses the same pixel coordinates on each image, slightly shifted star positions, larger FWHM, and PSF elongation result in the loss of counts in the aperture. This effect is slightly brightness-dependent as bright stars, which have more counts above the background noise level in the wings of the PSF, are less affected than faint stars.

\end{itemize}

\begin{table}
	\centering
		\begin{tabular}{rcccc}
			\hline
Range [1] & $r'$ [2] & $r'$ [3] & $i'$ [2] & $i'$ [3] \\
\hline

7	-- ~~8	&		&		&0.007	&0.005\\
8	-- ~~9	&0.007	&0.004	&0.009	&0.005\\
9	-- 10	&0.009	&0.004	&0.010	&0.006\\
10	-- 11	&0.007	&0.005	&0.011	&0.006\\
11	-- 12	&0.009	&0.005	&0.012	&0.007\\
12	-- 13	&0.011	&0.006	&0.018	&0.010\\
13	-- 14	&0.017	&0.008	&0.033	&0.015\\
14	-- 15	&0.034	&0.015	&0.074	&0.034\\
15	-- 16	&0.082	&0.036	&0.171	&0.072\\
16	-- 17	&0.184	&0.081	&0.305	&0.162\\
17  -- 18   &0.295	&0.189	&		&	\\
			\hline
		\end{tabular}
	\caption[Calibration stability]{Calibration stability of all stars [2] and adopted most constant stars [3] within a magnitude range [1] as a function of brightness for field GDS~0644-0035. [2] and [3] are determined as the mean night-to-night magnitude difference within a light curve.}
	\label{tab:calib}
\end{table}

\section{The variable star catalog in separate filters}

\subsection{Identification of variable stars in separate filters}

As explained in Paper~I, we concatenate the results of three different methods for both filters separately, i.e., the $A$ method using each light curve's amplitude $A$, the SD method using its standard deviation, and the $J$ method derived from the Stetson-$J$-Index (Stetson 1996). In combination with our own flagging system and the information of both filters we create the final list of variable sources.\par
\begin{figure}
\includegraphics[width=\columnwidth]{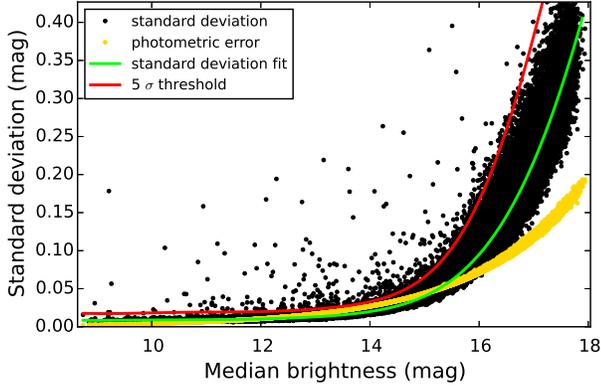}
\caption{Threshold and fit for SD~method, GDS 0700-0428 in $r'$}
\label{fig:sigma}
\end{figure}

To determine whether a source is variable we calculate a 5$\,\sigma$ threshold for both the $A$ and the SD method. The remaining average brightness fluctuations within a light curve after the ca\-li\-bra\-tion represent the intrinsic error still present in otherwise constant light curves. This effect results in a lowest detectable variation of truly variable sources, for both the amplitude and the light curve's standard deviation. It has to be determined as a function of brightness for both methods, since the scatter is highly brightness dependent.
The formal errors (including error propagation) of the individual data points of a light curve may be underestimated. Therefore, for comparison we give an additional error estimated from the global scatter of light curves of constant stars for different brightness ranges (Fig. \ref{fig:sigma}). These global scatter values are typically by a factor of 2 larger than the above formal errors for faint stars. Nevertheless, they do not affect the identification of variable stars with both the amplitude and sigma criteria methods, because these criteria use brightness dependent thresholds. Comparing light curves of stars in overlap regions of the fields shows that the resulting error between adjacent independently calibrated fields is on average considerably below $\pm$ 0.05\,mag in the image corner regions. One may expect that the uncertainty is even smaller in the image center regions.\par
To estimate the threshold for the SD method we fit the brightness-dependent standard deviations of all light curves with an $8^{th}$ degree polynomial $\sigma_{\mathrm{fit}}(m)$, representing this function very well for the entire magnitude range. It is a compromise between fit stability and adaptiveness to the data, which was not achieved as accurately with multiple linear fits or exponential functions. We then apply a 5$\,\sigma$ threshold to this fit, shown in Figure \ref{fig:sigma}.

\begin{figure}
\includegraphics[width=\columnwidth]{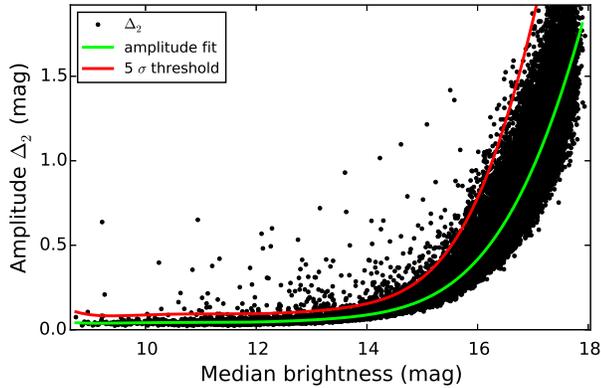}
\caption{Threshold and fit for $A$~method, GDS 0700-0428 in $r'$}
\label{fig:amplitude}
\end{figure}

To assess the offset for the amplitude distribution we calculate the second highest amplitude $A_2$ for each star; the full amplitude is more prone to be influenced by possible single outliers like planet or asteroid transits within our solar system crossing the aperture. Using smaller amplitudes than  $A_2$ would mean to eliminate real but rare variable events, thus we chose $A_2$ as a reasonable trade-off between robustness and sensitivity. We fit this amplitude as a function of the median light curve magnitude $m$ in the same way as for the standard deviation yielding an amplitude function $a(m)$ that allows us to use an amplitude cutoff based on each star's individual average brightness. We also apply a 5$\,\sigma$ threshold as described for the SD method; this is shown in Figure~\ref{fig:amplitude}.

\begin{figure}[h!]
\includegraphics[width=\columnwidth]{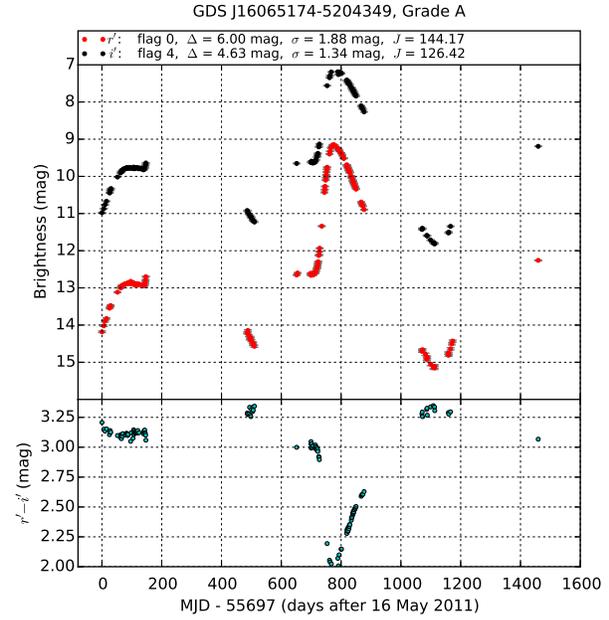}
\caption{Light curves and color curve for V\,352~Nor. The large amplitude of the variation causes sporadic saturation in $i'$, which, however, does not result in an automatic rejection of the light curve.}
\label{fig:mira}
\end{figure}

The criteria for a source to be identified as variable are as follows:

\begin{itemize}
\item [--]
The standard deviation $\sigma_{\mathrm{LC}}$ of a light curve exceeds the fit value of $\sigma_{\mathrm{fit}}(m)$ for its median brightness $m$ by $5\,\sigma_{\mathrm{SD}}$.

\begin{equation}
\sigma_{\mathrm{LC}} > \sigma_{\mathrm{fit}}(m) + 5 \,\sigma_{\mathrm{SD}}
\label{eq:sigma}
\end{equation}

\item [--]
The second highest amplitude (and therefore also the full amplitude) within a light curve has to exceed the corresponding fit value $a(m)$ by 5$\,\sigma_A$.

\begin{equation}
 A > A_{2} > a(m) + 5\,\sigma_A
\end{equation}

By that measure we get very few spurious variable detections. Also, we minimize the influence of single outliers while being still sensitive for light curves with only very few values deviating from the median brightness (e.g., eclipsing binaries with longer periods).

\begin{figure}[ht]
\includegraphics[width=\columnwidth]{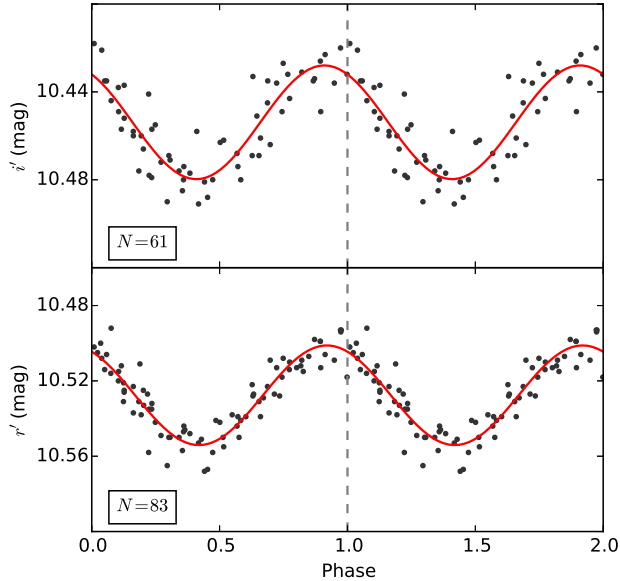}
\caption{HD~297501, a new periodic variable with $P$ = 0.6798\,\mbox{d}, and the smallest bonafide scatter-corrected amplitude of 0.053\,\mbox{mag} in $r'$ and 0.052\,\mbox{mag} in $i'$, respectively, taken from a sinusoidal least-squares fit (red line). The maximum amplitudes are $A$ = 0.076\,mag in $r'$ and $A$ = 0.073\,mag in $i'$.}
\label{fig:smallest}
\end{figure}

\item [--]
Using Stetson's $J$-Index, pairing subsequent measurements in each time series and using a threshold of 0.5 (Fruth et al. 2012), we obtain a lot of variables with small amplitudes with only very few spurious detections as well. For each field and filter, the distribution of the $J$ values, log($N$) vs. $J$, has roughly a Gaussian shape with FWHM($J$)  typically about 0.4, hence $\sigma$ about 0.17. Superimposed, the distributions have a tail toward high $J$ values recognizable as a kink at about $J$\,=\,0.3 (roughly 2\,$\sigma$) and typical height at log($N$)\,/\,3, indicating that there may be  variable sources at $J$\,$>$\,0.3. Therefore, for accepting a source as variable,
we have chosen a more conservative threshold $J$\,=\,0.5 (roughly 3\,$\sigma$). This high threshold may also counter-balance cases where the error in the light curve is underestimated. However, with our time sampling of usually $\Delta t \ge 1\,\mathrm{d}$ we find light curves that show a period in the order of a few days, often letting the magnitude jump from minimum to maximum brightness from one measurement to the next. To include those sources, in which $J$\,$<$\,0, we use the absolute value of J with the criterion

\begin{equation}
 |J| > 0.5
\end{equation}

Additionally, we require a $J$-Index detection to also show an amplitude with at least $A_{2} > a(m) + \sigma_A$, since we find positive detections with lower amplitudes often to result from the remaining systematic errors.
\end{itemize}

To be considered as a variable candidate a star has to be identified by at least one method and at least $10\%$ of the data points in the light curve have to be unsaturated. The latter assures that sources which are constantly flagged as saturated and therefore show artificial va\-ri\-a\-bi\-li\-ty due to bad pho\-to\-me\-try are excluded, while high-amplitude variables like Mira-type or eruptive variables are not discarded completely if some of their measurements have been saturated. Stars that are already saturated in the master image are omitted beforehand. Figure~\ref{fig:mira} shows a typical GDS light curve of Mira class star V\,352~Nor. It illustrates the afore mentioned partial saturation in $i'$ denoted by flag 4 that does not affect the variability detection of the otherwise high grade light curve.\par
For evenly sampled data other techniques like the autocorrelation of the light curves may be used to identify variable sources (e.g., Rebull et al. 2014). However, our light curves are sampled fairly inhomogeneously and any kind of interpolation of gaps will introduce additional uncertainties; therefore we did not apply the autocorrelation technique here.

\begin{figure}
\includegraphics[width=\columnwidth]{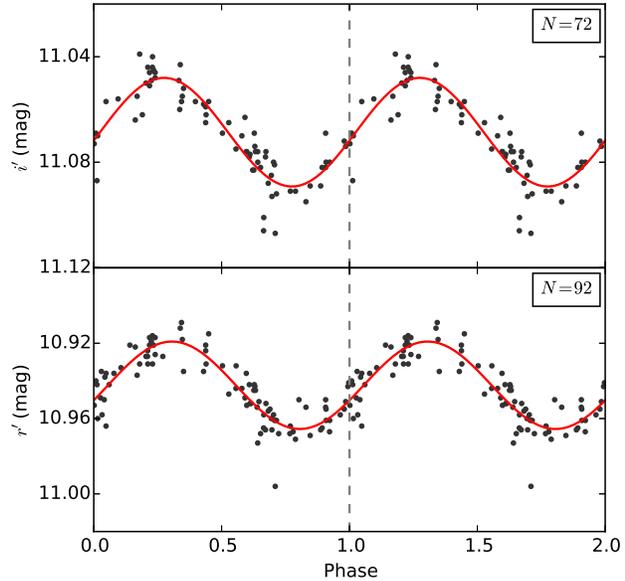}
\caption{V\,851~Mon, a known variable with $P$ = 0.3323\,\mbox{d} and a scatter-corrected amplitude of 0.046\,\mbox{mag} in $r'$ and 0.041\,\mbox{mag} in $i'$, respectively; VSX lists an amplitude of 0.05\,mag. The maximum amplitudes are $A$ = 0.087\,mag in $r'$ and $A$ = 0.068\,mag in $i'$.}
\label{fig:smallest_cat}
\end{figure}

\subsection{Filter combination and final variable detection}

To determine the final list of variables per field, we match the variables detected in both filters. Here we have to distinguish three different possible cases:

\begin{itemize}

\item [--] Case\,I: A source shows detectable variability in both filters

\item [--] Case\,II: A source is measured in both filters but exceeds the variability thresholds only in one filter

\item [--] Case\,III: A source is only detected in one filter

\end{itemize}

Additionally, we introduce a ranking system to assess the confidence in a positive detection using the advantage of simultaneous observation in two filters. This ranking system does not only represent the photometric quality of a light curve, but rather the confidence in a variability detection due to several different parameters. One confidence grade (A, B, or C) is assigned to each variable source based on the detection methods and the SE flags; grade A represents the highest confidence in a variability detection and grade C the lowest. The grades are based on which of the afore mentioned cases applies to a source in combination with the SE flags of both filters that represent the photometric quality and stability of a source's light curve. Grade A is assigned if a source is un-flagged in case\,I. Grade B is given if a source is uncritically flagged (i.e., SE flag $<$\,3) in case\,I or un-flagged in case\,II. Grade C is awarded to all sources in case\,I that are not classified as Grade A or B, to flagged case\,II sources with SE flag $<$\,3 and to all sources in case\,III. In case\,III, sources that show non-zero SE flags and/or are not detected by all three methods are considered very dubious and are therefore omitted for the final catalog of variable sources.

Due to the relatively sparse phase coverage in many fields, period search algorithms like the analysis of variance (AoV, Schwarzenberg-Czerny 1986, 1999) or Lomb-Scargle (LS, Lomb 1976; Scargle 1982) yielded highly irregular and faulty results on most of our light curves. Therefore, an automated period detection and classification approach has not yet been successful due to the very high number of false positives produced by those algorithms. Our AoV-based approach on period search and automated classification is under development and will be presented by Fein et al. (in prep.). However, using the Lafler-Kinman algorithm (see Paper~I, Lafler \& Kinman 1965) we have already been able to identify periods for selected stars in a period range of 5\,h up to $>$\,1000\,d for eclipsing and pulsating sources.

\begin{figure}
\includegraphics[width=\columnwidth]{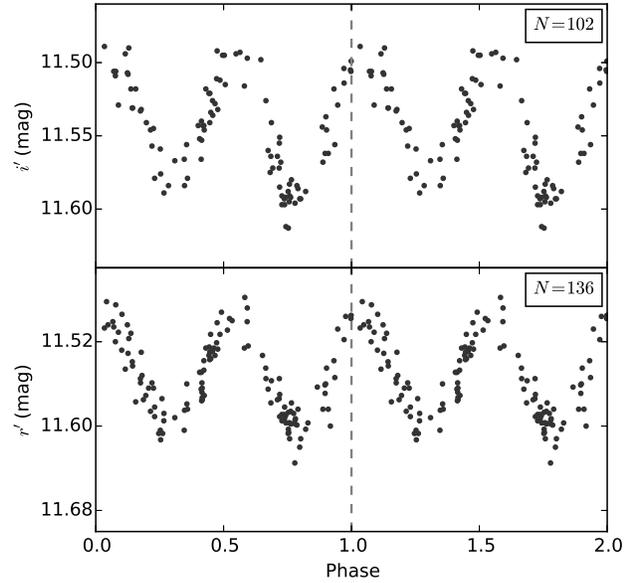}
\caption{TYC~7687-388-1, a suspected EW (W UMa) binary with the smallest automatically detected period ($P$ = 0.1996\,\mbox{d}) and a maximum amplitude of $A$ = 0.157\,\mbox{mag} in $r'$ and $A$ = 0.124\,\mbox{mag} in $i'$.}
\label{fig:shortest}
\end{figure}

To demonstrate the range and performance of our data we show a selection of preliminary results extensively discussed by Fein et al. (in prep.). Variable sources in the GDS get assigned a unique identifier of the form ``GDS Jhhmmsss\allowbreak $\pm$ddmmss'' of right ascension and declination, where J indicates coordinates for J2000. Figure~\ref{fig:smallest} shows the currently smallest amplitude of an automatically folded GDS light curve of HD~297501 which has not been previously classified as variable. In comparison, we were able to precisely verify the period of V\,851~Mon; its amplitude is listed by VSX as 0.05 mag which agrees well with our estimate of about 0.04 mag (Fig.~\ref{fig:smallest_cat}). The smallest period we were able to automatically detect yet is that of TYC~7687-388-1 (Fig.~\ref{fig:shortest}), which we suspect to be an eclipsing binary star with a period of 0.1998\,d. All periods are calculated as the average of the two independently detected periods in $r'$ and $i'$ with the criterion $| P_{r'} - P_{i'} | < 0.01\,\mbox{d}$. For the periods shown in Figures \ref{fig:smallest} -- \ref{fig:shortest} the difference between the two filters is $\le 0.00003\,\mbox{d}$.

\begin{figure}
\includegraphics[width=\columnwidth]{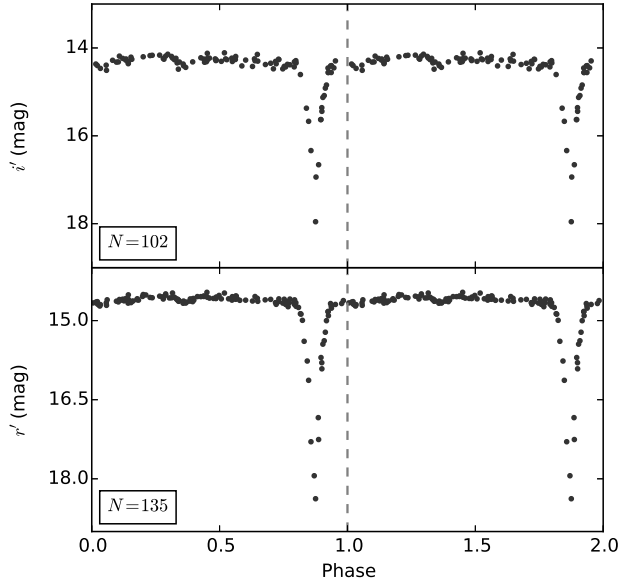}
\caption{GDS\,J0853028-443725, a suspected EA (Algol-type) binary with $P$ = 1.6726\,\mbox{d} and the currently largest binary amplitude of $A$ = 3.92\,\mbox{mag} in $r'$ and $A$ = 3.84\,\mbox{mag} in $i'$; the secondary minimum is extremely small.}
\label{fig:largest_eb}
\end{figure}

While we find copious numbers of long-term high-am\-pli\-tu\-de variables, the currently largest amplitude of an eclipsing binary with nearly $4\,\mbox{mag}$ and an extraordinarily small secondary minimum is that of GDS\,J0853028\allowbreak -443726 (Fig. \ref{fig:largest_eb}), which was not identified and folded automatically, but was discovered due to its high amplitude.

\section{The extended variable star catalog}
\label{sec_catalog}

We have analyzed 268 fields observed between September 2010 and May 2015 with com\-plete\-ness limits of $r'$\,$\sim$\,18\,\mbox{mag} and $i'$\,$\sim$\,17\,\mbox{mag}. All detected variable sources were matched with the VSX within a $10''$ radius; the VSX contains the major catalogs of variable stars including ASAS used in Paper~I.
Our sources have an average positional accuracy of 1'', and $\sim$\,99\% of our variable sources are separated by more than 20''. The ASAS coordinates are given with a precision of 0.1' (6'') in declination and 1$^{s}$ (up to 15'' at Dec\,=\,0$^{\circ}$ in right ascension. To match our sources with the ASAS catalog in Paper~I, we used a rectangular box
of 7'' in Dec and 16''\,/\,cos\,(Dec) in RA. The VSX coordinates are from different observers and instruments. To match our sources with VSX we used a -- quite large -- matching
radius of 10''. If more than one variable GDS source matches a VSX or ASAS source, we accepted the closer one. While the use of the large matching radius may yield false matches, it avoids that we overlook a former detection of a variable source. Thus, the number of new variables found in the GDS may be a bit larger.\par
The resulting catalog of variable stars contains the following information for the light curves of 64\,151 variable sources for $r'$ and $i'$ filters with amplitudes between $0.03$ and $7.5\,\mbox{mag}$.

 \begin{itemize}
\item [--] GDS Field
\item [--] RA mean (deg)
\item [--] Dec mean (deg)
\item [--] RA $r'$ (deg)
\item [--] Dec $r'$ (deg)
\item [--] RA $i'$ (deg)
\item [--] Dec $i'$ (deg)
\item [--] GDS Grade
\item [--] No. of Measurements $r'$
\item [--] No. of Measurements $i'$
\item [--] Median Magnitude $r'$
\item [--] Equivalent Average Standard Deviation $r'$ (mag) [1]
\item [--] Median Absolute Deviation MAD $r'$ (mag)
\item [--] Median Magnitude $i'$
\item [--] Equivalent Average Standard Deviation $i'$ (mag) [1]
\item [--] Median Absolute Deviation MAD $i'$ (mag)
\item [--] SE Flag $r'$
\item [--] SE Flag $i'$
\item [--] Maximum Amplitude (mag) [2]
\item [--] Amplitude Flag $r'$ [3]
\item [--] Amplitude Flag $i'$ [3]
\item [--] Amplitude $r'$ (mag)
\item [--] Amplitude $i'$ (mag)
\item [--] Amplitude Threshold $r'$ (mag)
\item [--] Amplitude Threshold $i'$ (mag)
\item [--] Maximum Stetson-$J$ [2]
\item [--] Stetson-$J$ Flag $r'$ [3]
\item [--] Stetson-$J$ Flag $i'$ [3]
\item [--] Stetson-$J$ $r'$
\item [--] Stetson-$J$ $i'$
\item [--] Stetson-$J$ Threshold $r'$
\item [--] Stetson-$J$ Threshold $i'$
\item [--] Maximum Standard Deviation (mag) [2]
\item [--] Standard Deviation Flag $r'$ [3]
\item [--] Standard Deviation Flag $i'$ [3]
\item [--] Standard Deviation $r'$ (mag)
\item [--] Standard Deviation $i'$ (mag)
\item [--] Standard Deviation Threshold $r'$ (mag)
\item [--] Standard Deviation Threshold $i'$ (mag)
\item [--] VSX match [4]
\item [--] VSX OID
 \end{itemize}

[1] is the value of $\sigma_{\mathrm{fit}}(m)$ from Equation \ref{eq:sigma}, [2] is the maximum value of both filters. [3] is 0 if a method was negative and 1 if positive. [4] is 1 if a match is found within VSX, 0 otherwise.
The full catalog of variable sources is available electronically or can be obtained from the authors. Additionally, we provide all light curves in a separate catalog listed in Table \ref{tab:lightcurves}.
Comparing the GDS variables catalog with VSX we find that 6823 sources are listed as variables and another 534 as candidate variables. That yields 56\,794 new variable sources within the GDS not listed in VSX.

\begin{figure}
\includegraphics[width=\columnwidth]{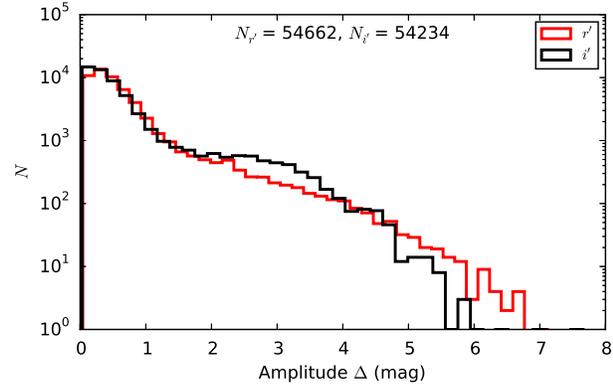}
\caption{Light curve amplitude distribution of variable sources detected by amplitude in $r'$ and $i'$}
\label{fig:amplitudes}
\end{figure}

\begin{figure}
\includegraphics[width=\columnwidth]{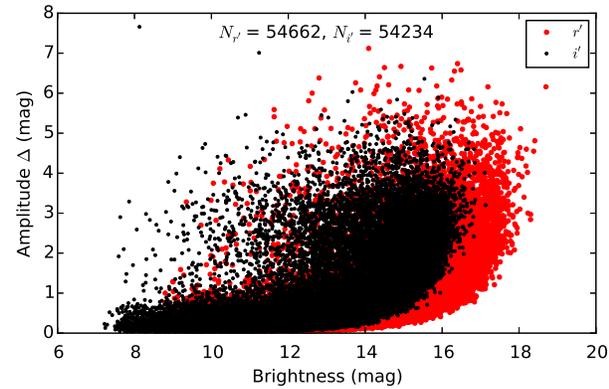}
\caption{Brightness-dependency of amplitudes in $r'$ and $i'$}
\label{fig:amplitudes2}
\end{figure}

The light curve amplitude histogram for all variable sour\-ces detected by amplitude is depicted in Figure~\ref{fig:amplitudes} and shows a slight abundance of sources between 2 and 4\,mag amplitudes in $i'$. This seems to be caused by red Mira-class variables that are too faint to be decently measured in $r'$. The corresponding brightness-dependency of those measured light curve amplitudes is shown in Figure~\ref{fig:amplitudes2}.

\begin{table*}
	\centering
		\begin{tabular}{cccccccc}
			\hline

GDS Field & RA (J2000)  & Dec (J2000) & GDS ID & MJD & Filter &  Magnitude & Error \\
\hline

GDS\_0644-0035&100.05585&-1.24677&GDS\_J0640134-011448&55945.11762&$r'$&13.0389&0.0124\\
GDS\_0644-0035&100.05585&-1.24677&GDS\_J0640134-011448&55951.11568&$r'$&13.0409&0.0124\\
GDS\_0644-0035&100.05585&-1.24677&GDS\_J0640134-011448&55961.19869&$r'$&13.1529&0.0129\\
GDS\_0644-0035&100.05585&-1.24677&GDS\_J0640134-011448&56184.36252&$r'$&13.1056&0.0127\\
GDS\_0644-0035&100.05585&-1.24677&GDS\_J0640134-011448&56191.33197&$r'$&13.0900&0.0126\\
GDS\_0644-0035&100.05585&-1.24677&GDS\_J0640134-011448&55945.11762&$i'$&12.9079&0.0187\\
GDS\_0644-0035&100.05585&-1.24677&GDS\_J0640134-011448&55961.19869&$i'$&13.0023&0.0198\\
GDS\_0644-0035&100.05585&-1.24677&GDS\_J0640134-011448&56184.36252&$i'$&12.9393&0.0190\\
GDS\_0644-0035&100.05585&-1.24677&GDS\_J0640134-011448&56191.33197&$i'$&12.9293&0.0189\\
GDS\_0644-0035&100.05585&-1.24677&GDS\_J0640134-011448&56201.31892&$i'$&12.9191&0.0188\\

			\hline
		\end{tabular}
	\caption[Light curve catalog]{Light curve catalog of all variable GDS sources. For illustration only the entries for one source in both filters are printed. The full catalog is available electronically or can be obtained from the authors.}
	\label{tab:lightcurves}
\end{table*}

\begin{figure}
\includegraphics[width=\columnwidth]{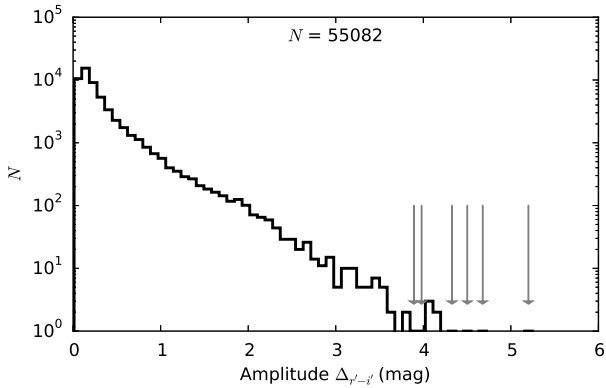}
\caption{$r'\!-\!i'$ color amplitude distribution for all variable sources measured in both filters; arrows denote bins with only one object.}
\label{fig:colors}
\end{figure}

Additionally, we compute $r'\!-\!i'$ for each pair of observations within a light curve for each source measured in both filters. This yields a color curve, which documents the change in color for each source in addition to its $r'$ and $i'$ light curves. The histogram of color curve amplitudes is shown in Figure~\ref{fig:colors}.

\section{The multi-wavelength photometric catalog}

\begin{figure}
\includegraphics[width=\columnwidth]{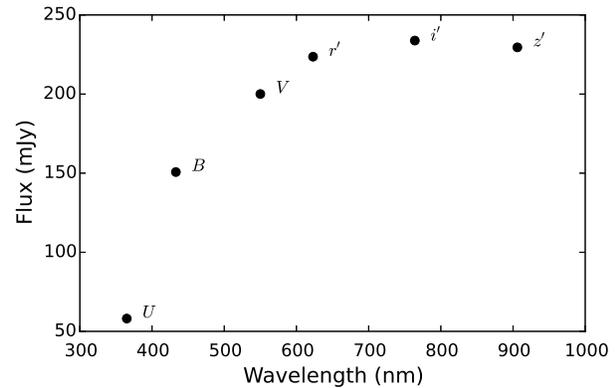}
\caption{Spectral energy distribution of HD\,297501 measured in all six filters}
\label{fig:sed}
\end{figure}

Additionally to the variability monitoring, we have extended our observation scheme to provide photometry for the filters $U$, $B$, $V$, $r'$, $i'$ and $z'$. It is the aim to observe each field at least three times in the extended filter set to obtain a rough spectral energy distribution based on six wavelengths. This multi-color catalog stores the magnitudes for 15\,888\,583 sources together with its SE flags. For that we determine the median brightness in each filter to get a more stable result as compared to the average (especially in case of variable sources). Hence, we compute each source's median absolute deviation (MAD) as an error measure, which is also given in the catalog. To assess the constancy of a light curve -- and therefore the confidence in the accuracy of a given median brightness -- we also store the magnitude amplitude of the available data points in each filter. Each source in this constantly updated catalog has at least a magnitude for one of the six filters. The catalog is shown in Table \ref{tab:photfilters}. Exemplary, Figure~\ref{fig:sed} shows the measured SED of HD\,297501 (compare Figure~\ref{fig:smallest}).

\begin{table*}
	\centering
		\begin{tabular}{cccccccr}
			\hline

RA (J2000)  & Dec (J2000) & Filter &  Magnitude [1] & Error [2] & Amplitude [3] & SE Flag [4] & $N_{obs}$ [5]\\
\hline

100.04751&-3.03723&$U$&13.896&0.020&0.062&0&4\\
100.04751&-3.03723&$B$&13.679&0.004&0.009&0&5\\
100.04751&-3.03723&$V$&13.124&0.004&0.014&0&4\\
100.04751&-3.03723&$r'$&12.956&0.007&0.056&0&114\\
100.04751&-3.03723&$i'$&12.854&0.015&0.103&0&101\\
100.04751&-3.03723&$z'$&12.771&0.020&0.064&0&4\\

			\hline
		\end{tabular}
	\caption[Photometric catalog]{Photometric catalog of all GDS sources. [1] is the median magnitude of the available measurements, [2] is the MAD and [3] the full amplitude computed from those measurements. [4] is the photometry flag given by SE and [5] is the number of available observations per source and filter. For illustration only the entries for one source are printed. The full catalog is available electronically or can be obtained from the authors.}
	\label{tab:photfilters}
\end{table*}

\section{Outlook}

All GDS observations are on-going to further extend the time and phase coverage and we are updating the release on a regular basis. The catalogs are available electronically and can be requested from the authors.\par
While VPHAS+ (although without variability information) complements the GDS with respect to close and faint sources, the conjunction of VVV (albeit regrettably not further continued beyond 2016 in contrast to our survey) and GDS makes it possible to link the extensive variability studies from optical to infrared.

\acknowledgements{%
This work was supported by the Nordrhein--Westf\"{a}lische Akademie der Wissenschaften und der K\"{u}nste, funded by the Federal State Nordrhein--Westfalen and the Federal Republic of Germany.
The observations on Cerro Armazones benefitted from the care of the guards Hector Labra, Gerard Pino, Alberto Lavin, and Francisco Arraya and from the constant support of the Universidad Cat\'{o}lica del Norte. 
Special thanks go to Holger Drass, Roland Lemke, and Ramon Watermann, for their invaluable technical support. This research has made use of the International Variable Star Index (VSX) database, operated at AAVSO, Cambridge, Massachusetts, USA.

}

\appendix
\section{Coverage}

\begin{figure*}
\includegraphics[width=\textwidth]{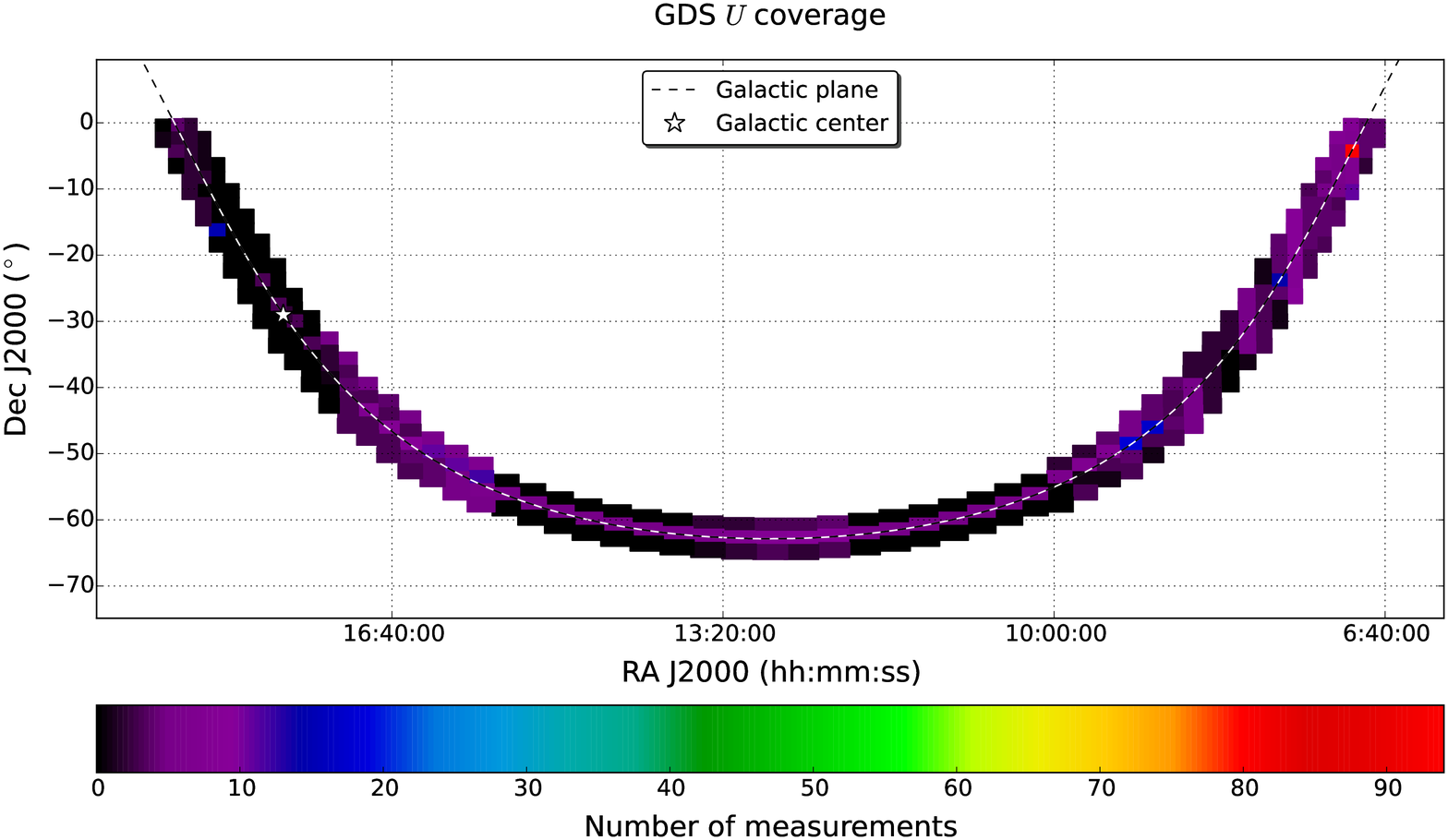}
\caption{$U$ field coverage}
\label{fig:U_coverage}
\end{figure*}

\begin{figure*}
\includegraphics[width=\textwidth]{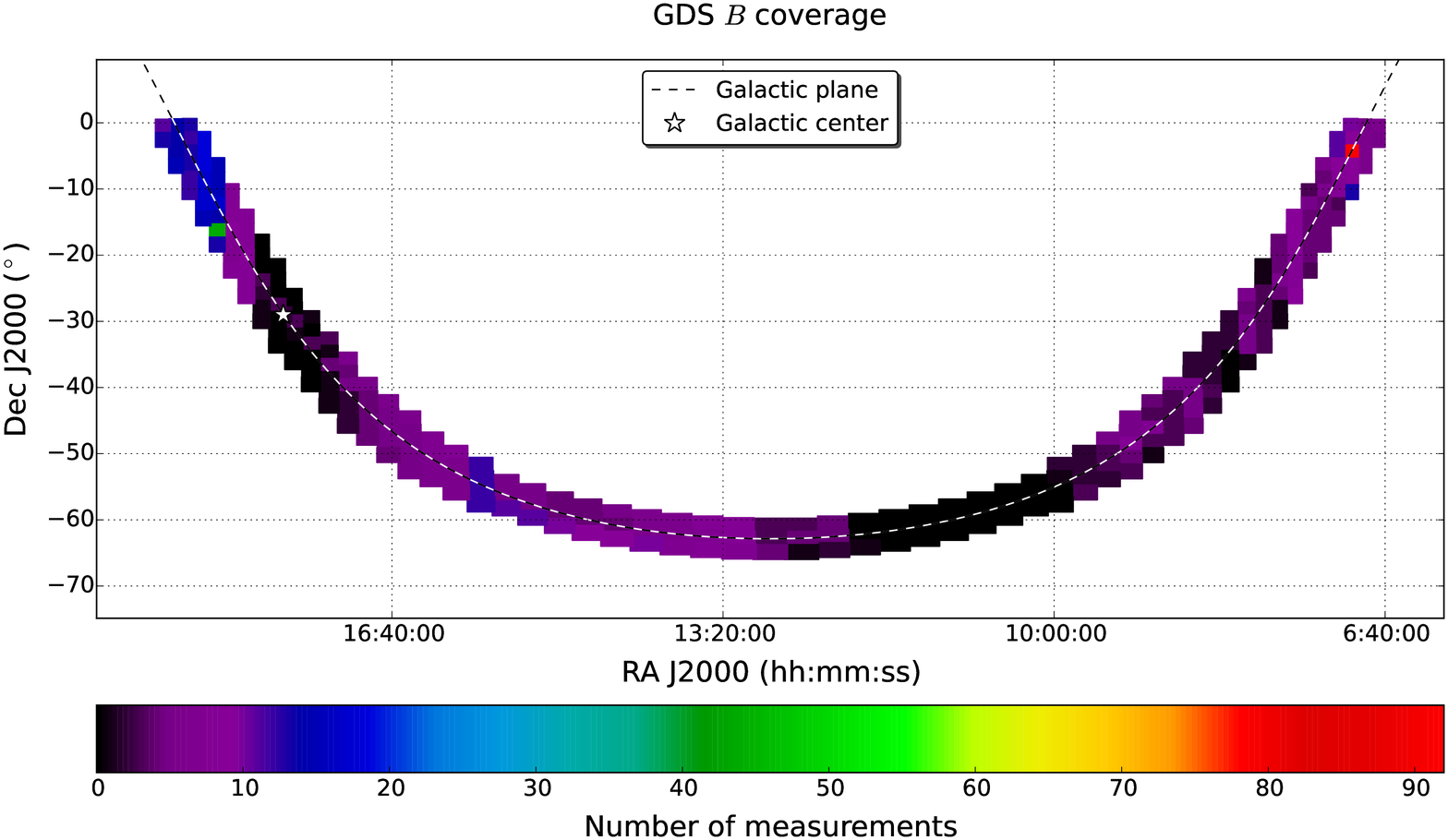}
\caption{$B$ field coverage}
\label{fig:B_coverage}
\end{figure*}

\begin{figure*}
\includegraphics[width=\textwidth]{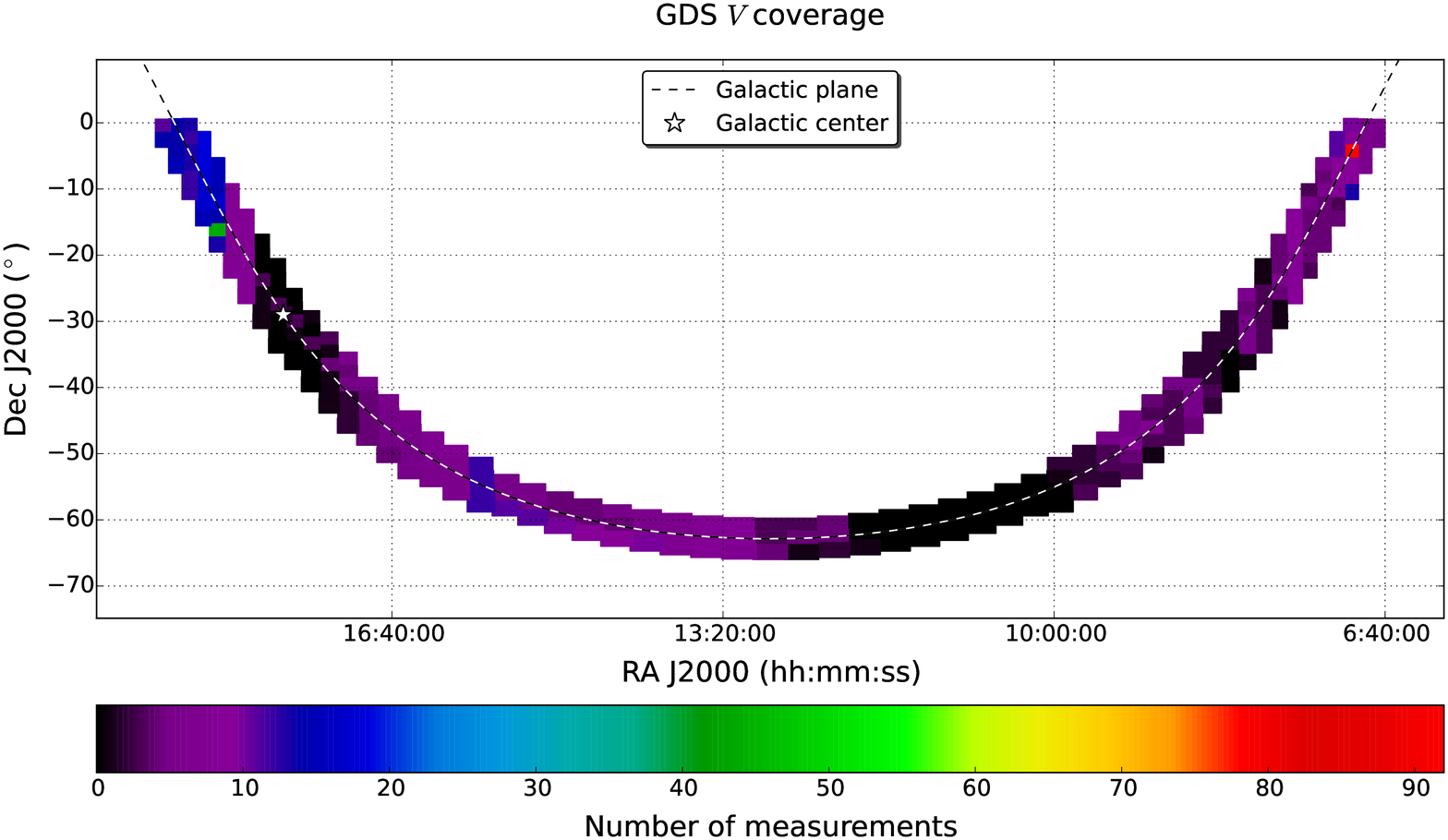}
\caption{$V$ field coverage}
\label{fig:V_coverage}
\end{figure*}

\begin{figure*}
\includegraphics[width=\textwidth]{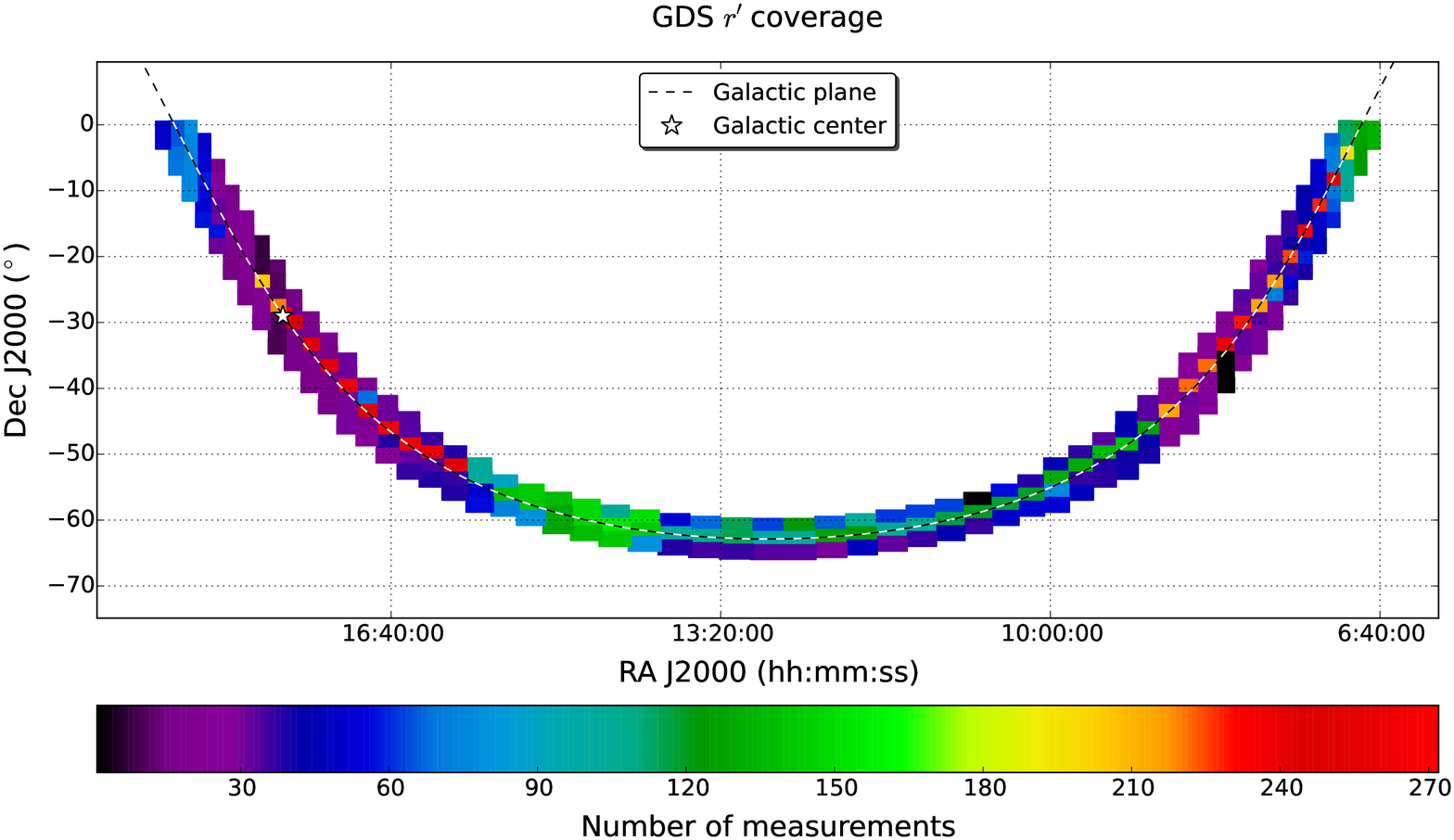}
\caption{$r'$ field coverage}
\label{fig:r_coverage}
\end{figure*}

\begin{figure*}
\includegraphics[width=\textwidth]{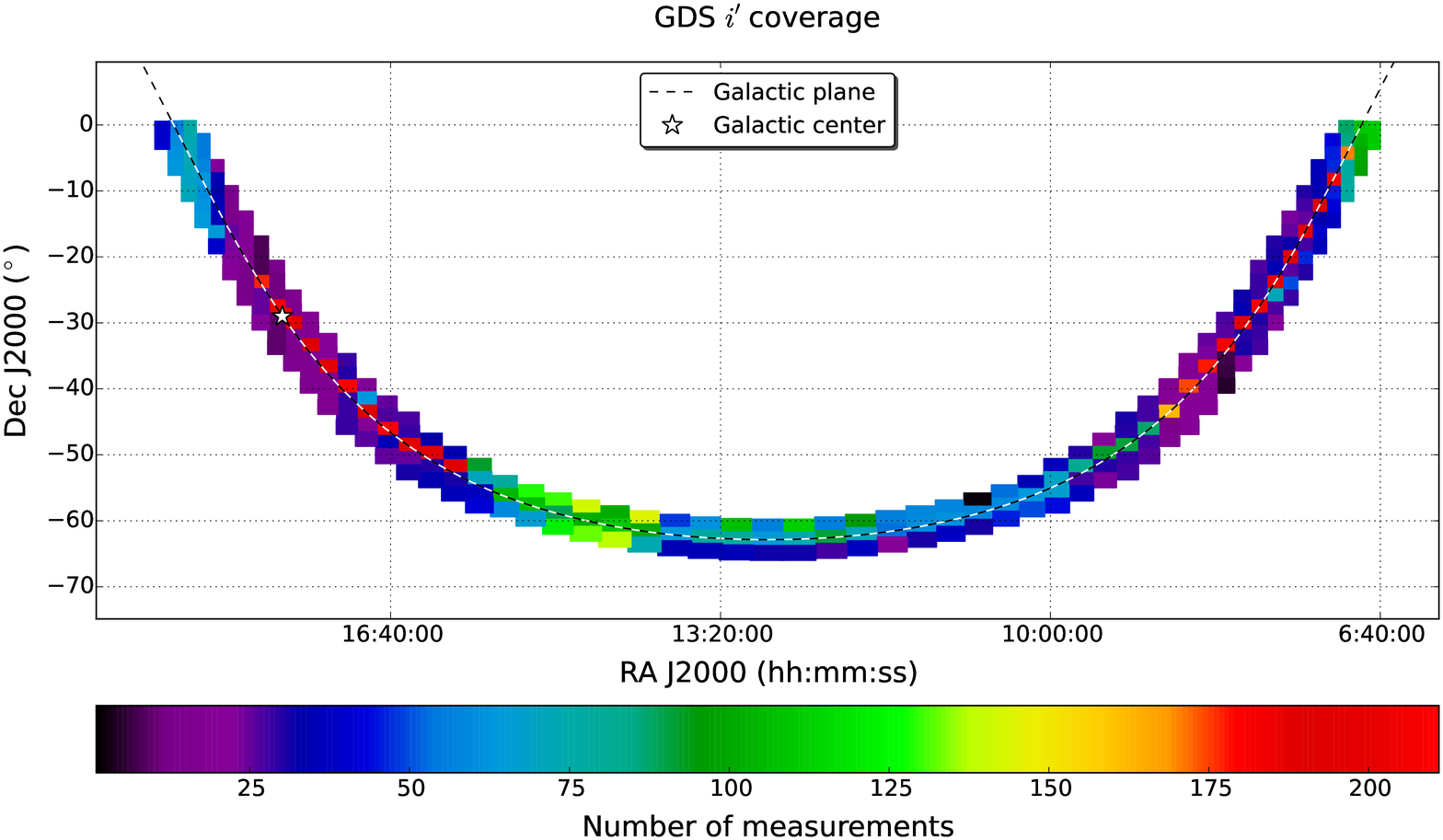}
\caption{$i'$ field coverage}
\label{fig:i_coverage}
\end{figure*}

\begin{figure*}
\includegraphics[width=\textwidth]{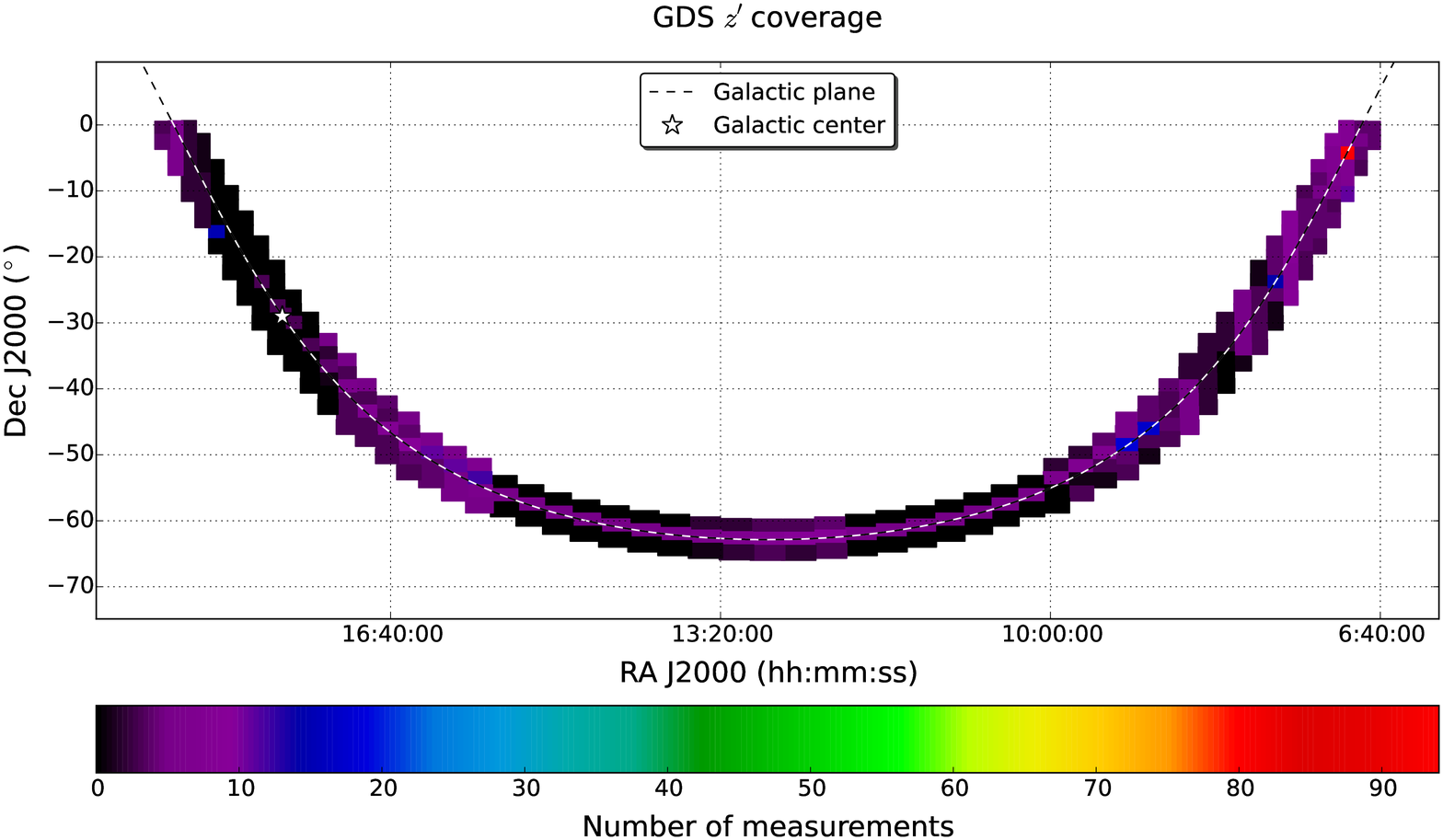}
\caption{$z'$ field coverage}
\label{fig:z_coverage}
\end{figure*}

\renewcommand{\arraystretch}{1.25}
\setlength{\tabcolsep}{2pt}
\begin{table*}
\fontsize{7}{7}\selectfont
	\centering
		\begin{tabular}{ccc @{\hspace{2em}}ccc @{\hspace{2em}}ccc @{\hspace{2em}}ccc}
			\hline
GDS field & RA [1] & Dec [2] & GDS field & RA [1] & Dec [2] & GDS field & RA [1] & Dec [2] & GDS field & RA [1] & Dec [2] \\
\hline
0644-0035 & 06:44:27.5 & -01:00:35 & 0824-3445 & 08:24:55.8 & -01:34:45 & 1251-6452 & 12:51:30.5 & -01:64:52 & 1728-2931 & 17:28:53.3 & -01:29:31\\
0644-0235 & 06:44:27.5 & -01:02:35 & 0824-3645 & 08:24:55.8 & -01:36:45 & 1310-6047 & 13:10:20.9 & -01:60:47 & 1728-3131 & 17:28:53.3 & -01:31:31\\
0652-0031 & 06:52:30.6 & -01:00:31 & 0824-3845 & 08:24:55.8 & -01:38:45 & 1310-6247 & 13:10:20.9 & -01:62:47 & 1728-3331 & 17:28:53.3 & -01:33:31\\
0652-0231 & 06:52:30.6 & -01:02:31 & 0824-4045 & 08:24:55.8 & -01:40:45 & 1310-6447 & 13:10:20.9 & -01:64:47 & 1728-3531 & 17:28:53.3 & -01:35:31\\
0652-0431 & 06:52:30.6 & -01:04:31 & 0824-4245 & 08:24:55.8 & -01:42:45 & 1329-6033 & 13:29:07.9 & -01:60:33 & 1728-3731 & 17:28:53.3 & -01:37:31\\
0652-0631 & 06:52:30.6 & -01:06:31 & 0836-3547 & 08:36:24.2 & -01:35:47 & 1329-6233 & 13:29:07.9 & -01:62:33 & 1728-3931 & 17:28:53.3 & -01:39:31\\
0700-0028 & 07:00:38.7 & -01:00:28 & 0836-3747 & 08:36:24.2 & -01:37:47 & 1329-6433 & 13:29:07.9 & -01:64:33 & 1739-2609 & 17:39:15.6 & -01:26:09\\
0700-0228 & 07:00:38.7 & -01:02:28 & 0836-3947 & 08:36:24.2 & -01:39:47 & 1347-6009 & 13:47:45.1 & -01:60:09 & 1739-2809 & 17:39:15.6 & -01:28:09\\
0700-0428 & 07:00:38.7 & -01:04:28 & 0836-4147 & 08:36:24.2 & -01:41:47 & 1347-6209 & 13:47:45.1 & -01:62:09 & 1739-3009 & 17:39:15.6 & -01:30:09\\
0700-0628 & 07:00:38.7 & -01:06:28 & 0836-4347 & 08:36:24.2 & -01:43:47 & 1347-6409 & 13:47:45.1 & -01:64:09 & 1739-3209 & 17:39:15.6 & -01:32:09\\
0700-0828 & 07:00:38.7 & -01:08:28 & 0836-4547 & 08:36:24.2 & -01:45:47 & 1406-5935 & 14:06:06.2 & -01:59:35 & 1739-3409 & 17:39:15.6 & -01:34:09\\
0700-1028 & 07:00:38.7 & -01:10:28 & 0848-3936 & 08:48:16.2 & -01:39:36 & 1406-6135 & 14:06:06.2 & -01:61:35 & 1739-3609 & 17:39:15.6 & -01:36:09\\
0708-0227 & 07:08:54.4 & -01:02:27 & 0848-4136 & 08:48:16.2 & -01:41:36 & 1406-6335 & 14:06:06.2 & -01:63:35 & 1745-2856 & 17:45:37.2 & -01:28:56\\
0708-0427 & 07:08:54.4 & -01:04:27 & 0848-4336 & 08:48:16.2 & -01:43:36 & 1424-5852 & 14:24:05.6 & -01:58:52 & 1749-2139 & 17:49:10.1 & -01:21:39\\
0708-0627 & 07:08:54.4 & -01:06:27 & 0848-4536 & 08:48:16.2 & -01:45:36 & 1424-6052 & 14:24:05.6 & -01:60:52 & 1749-2339 & 17:49:10.1 & -01:23:39\\
0708-0827 & 07:08:54.4 & -01:08:27 & 0848-4736 & 08:48:16.2 & -01:47:36 & 1424-6252 & 14:24:05.6 & -01:62:52 & 1749-2539 & 17:49:10.1 & -01:25:39\\
0708-1027 & 07:08:54.4 & -01:10:27 & 0900-4213 & 09:00:46.5 & -01:42:13 & 1441-5758 & 14:41:38.2 & -01:57:58 & 1749-2739 & 17:49:10.1 & -01:27:39\\
0708-1227 & 07:08:54.4 & -01:12:27 & 0900-4413 & 09:00:46.5 & -01:44:13 & 1441-5958 & 14:41:38.2 & -01:59:58 & 1749-2939 & 17:49:10.1 & -01:29:39\\
0708-1427 & 07:08:54.4 & -01:14:27 & 0900-4613 & 09:00:46.5 & -01:46:13 & 1441-6158 & 14:41:38.2 & -01:61:58 & 1749-3139 & 17:49:10.1 & -01:31:39\\
0717-0624 & 07:17:20.3 & -01:06:24 & 0900-4813 & 09:00:46.5 & -01:48:13 & 1458-5654 & 14:58:39.7 & -01:56:54 & 1749-3339 & 17:49:10.1 & -01:33:39\\
0717-0824 & 07:17:20.3 & -01:08:24 & 0900-5013 & 09:00:46.5 & -01:50:13 & 1458-5854 & 14:58:39.7 & -01:58:54 & 1758-1757 & 17:58:46.7 & -01:17:57\\
0717-1024 & 07:17:20.3 & -01:10:24 & 0913-4440 & 09:13:58.2 & -01:44:40 & 1458-6054 & 14:58:39.7 & -01:60:54 & 1758-1957 & 17:58:46.7 & -01:19:57\\
0717-1224 & 07:17:20.3 & -01:12:24 & 0913-4640 & 09:13:58.2 & -01:46:40 & 1515-5540 & 15:15:07.2 & -01:55:40 & 1758-2157 & 17:58:46.7 & -01:21:57\\
0717-1424 & 07:17:20.3 & -01:14:24 & 0913-4840 & 09:13:58.2 & -01:48:40 & 1515-5740 & 15:15:07.2 & -01:57:40 & 1758-2357 & 17:58:46.7 & -01:23:57\\
0717-1624 & 07:17:20.3 & -01:16:24 & 0913-5040 & 09:13:58.2 & -01:50:40 & 1515-5940 & 15:15:07.2 & -01:59:40 & 1758-2557 & 17:58:46.7 & -01:25:57\\
0717-1824 & 07:17:20.3 & -01:18:24 & 0913-5240 & 09:13:58.2 & -01:52:40 & 1530-5416 & 15:30:57.9 & -01:54:16 & 1758-2757 & 17:58:46.7 & -01:27:57\\
0725-1020 & 07:25:59.2 & -01:10:20 & 0927-4752 & 09:27:32.3 & -01:47:52 & 1530-5616 & 15:30:57.9 & -01:56:16 & 1758-2957 & 17:58:46.7 & -01:29:57\\
0725-1220 & 07:25:59.2 & -01:12:20 & 0927-4952 & 09:27:32.3 & -01:49:52 & 1530-5816 & 15:30:57.9 & -01:58:16 & 1808-1411 & 18:08:00.8 & -01:14:11\\
0725-1420 & 07:25:59.2 & -01:14:20 & 0927-5152 & 09:27:32.3 & -01:51:52 & 1546-5141 & 15:46:10.7 & -01:51:41 & 1808-1611 & 18:08:00.8 & -01:16:11\\
0725-1620 & 07:25:59.2 & -01:16:20 & 0927-5352 & 09:27:32.3 & -01:53:52 & 1546-5341 & 15:46:10.7 & -01:53:41 & 1808-1811 & 18:08:00.8 & -01:18:11\\
0725-1820 & 07:25:59.2 & -01:18:20 & 0941-4951 & 09:41:47.7 & -01:49:51 & 1546-5541 & 15:46:10.7 & -01:55:41 & 1808-2011 & 18:08:00.8 & -01:20:11\\
0725-2020 & 07:25:59.2 & -01:20:20 & 0941-5151 & 09:41:47.7 & -01:51:51 & 1546-5741 & 15:46:10.7 & -01:57:41 & 1808-2211 & 18:08:00.8 & -01:22:11\\
0725-2220 & 07:25:59.2 & -01:22:20 & 0941-5351 & 09:41:47.7 & -01:53:51 & 1601-4952 & 16:01:08.7 & -01:49:52 & 1808-2411 & 18:08:00.8 & -01:24:11\\
0734-1411 & 07:34:54.2 & -01:14:11 & 0941-5551 & 09:41:47.7 & -01:55:51 & 1601-5152 & 16:01:08.7 & -01:51:52 & 1808-2611 & 18:08:00.8 & -01:26:11\\
0734-1611 & 07:34:54.2 & -01:16:11 & 0956-5141 & 09:56:45.9 & -01:51:41 & 1601-5352 & 16:01:08.7 & -01:53:52 & 1816-1019 & 18:16:55.7 & -01:10:19\\
0734-1811 & 07:34:54.2 & -01:18:11 & 0956-5341 & 09:56:45.9 & -01:53:41 & 1601-5552 & 16:01:08.7 & -01:55:52 & 1816-1219 & 18:16:55.7 & -01:12:19\\
0734-2011 & 07:34:54.2 & -01:20:11 & 0956-5541 & 09:56:45.9 & -01:55:41 & 1615-4751 & 16:15:24.2 & -01:47:51 & 1816-1419 & 18:16:55.7 & -01:14:19\\
0734-2211 & 07:34:54.2 & -01:22:11 & 0956-5741 & 09:56:45.9 & -01:57:41 & 1615-4951 & 16:15:24.2 & -01:49:51 & 1816-1619 & 18:16:55.7 & -01:16:19\\
0734-2411 & 07:34:54.2 & -01:24:11 & 1011-5416 & 10:11:58.7 & -01:54:16 & 1615-5151 & 16:15:24.2 & -01:51:51 & 1816-1819 & 18:16:55.7 & -01:18:19\\
0734-2611 & 07:34:54.2 & -01:26:11 & 1011-5616 & 10:11:58.7 & -01:56:16 & 1615-5351 & 16:15:24.2 & -01:53:51 & 1816-2019 & 18:16:55.7 & -01:20:19\\
0744-1758 & 07:44:08.3 & -01:17:58 & 1011-5816 & 10:11:58.7 & -01:58:16 & 1628-4439 & 16:28:58.1 & -01:44:39 & 1816-2219 & 18:16:55.7 & -01:22:19\\
0744-1958 & 07:44:08.3 & -01:19:58 & 1027-5541 & 10:27:49.7 & -01:55:41 & 1628-4639 & 16:28:58.1 & -01:46:39 & 1825-0623 & 18:25:34.6 & -01:06:23\\
0744-2158 & 07:44:08.3 & -01:21:58 & 1027-5741 & 10:27:49.7 & -01:57:41 & 1628-4839 & 16:28:58.1 & -01:48:39 & 1825-0823 & 18:25:34.6 & -01:08:23\\
0744-2358 & 07:44:08.3 & -01:23:58 & 1027-5941 & 10:27:49.7 & -01:59:41 & 1628-5039 & 16:28:58.1 & -01:50:39 & 1825-1023 & 18:25:34.6 & -01:10:23\\
0744-2558 & 07:44:08.3 & -01:25:58 & 1044-5655 & 10:44:17.4 & -01:56:55 & 1628-5239 & 16:28:58.1 & -01:52:39 & 1825-1223 & 18:25:34.6 & -01:12:23\\
0744-2758 & 07:44:08.3 & -01:27:58 & 1044-5855 & 10:44:17.4 & -01:58:55 & 1642-4212 & 16:42:09.5 & -01:42:12 & 1825-1423 & 18:25:34.6 & -01:14:23\\
0744-2958 & 07:44:08.3 & -01:29:58 & 1044-6055 & 10:44:17.4 & -01:60:55 & 1642-4412 & 16:42:09.5 & -01:44:12 & 1825-1623 & 18:25:34.6 & -01:16:23\\
0753-2140 & 07:53:45.1 & -01:21:40 & 1101-5758 & 11:01:19.2 & -01:57:58 & 1642-4612 & 16:42:09.5 & -01:46:12 & 1825-1823 & 18:25:34.6 & -01:18:23\\
0753-2340 & 07:53:45.1 & -01:23:40 & 1101-5958 & 11:01:19.2 & -01:59:58 & 1642-4812 & 16:42:09.5 & -01:48:12 & 1834-0226 & 18:34:00.4 & -01:02:26\\
0753-2540 & 07:53:45.1 & -01:25:40 & 1101-6158 & 11:01:19.2 & -01:61:58 & 1642-5012 & 16:42:09.5 & -01:50:12 & 1834-0426 & 18:34:00.4 & -01:04:26\\
0753-2740 & 07:53:45.1 & -01:27:40 & 1118-5852 & 11:18:51.9 & -01:58:52 & 1654-3935 & 16:54:39.7 & -01:39:35 & 1834-0626 & 18:34:00.4 & -01:06:26\\
0753-2940 & 07:53:45.1 & -01:29:40 & 1118-6052 & 11:18:51.9 & -01:60:52 & 1654-4135 & 16:54:39.7 & -01:41:35 & 1834-0826 & 18:34:00.4 & -01:08:26\\
0753-3140 & 07:53:45.1 & -01:31:40 & 1118-6252 & 11:18:51.9 & -01:62:52 & 1654-4335 & 16:54:39.7 & -01:43:35 & 1834-1026 & 18:34:00.4 & -01:10:26\\
0753-3340 & 07:53:45.1 & -01:33:40 & 1136-5935 & 11:36:51.3 & -01:59:35 & 1654-4535 & 16:54:39.7 & -01:45:35 & 1834-1226 & 18:34:00.4 & -01:12:26\\
0803-2610 & 08:03:39.7 & -01:26:10 & 1136-6135 & 11:36:51.3 & -01:61:35 & 1654-4735 & 16:54:39.7 & -01:47:35 & 1834-1426 & 18:34:00.4 & -01:14:26\\
0803-2810 & 08:03:39.7 & -01:28:10 & 1136-6335 & 11:36:51.3 & -01:63:35 & 1706-3546 & 17:06:31.5 & -01:35:46 & 1842-0027 & 18:42:16.1 & -01:00:27\\
0803-3010 & 08:03:39.7 & -01:30:10 & 1155-6009 & 11:55:12.5 & -01:60:09 & 1706-3746 & 17:06:31.5 & -01:37:46 & 1842-0227 & 18:42:16.1 & -01:02:27\\
0803-3210 & 08:03:39.7 & -01:32:10 & 1155-6209 & 11:55:12.5 & -01:62:09 & 1706-3946 & 17:06:31.5 & -01:39:46 & 1842-0427 & 18:42:16.1 & -01:04:27\\
0803-3410 & 08:03:39.7 & -01:34:10 & 1155-6409 & 11:55:12.5 & -01:64:09 & 1706-4146 & 17:06:31.5 & -01:41:46 & 1842-0627 & 18:42:16.1 & -01:06:27\\
0803-3610 & 08:03:39.7 & -01:36:10 & 1213-6033 & 12:13:49.8 & -01:60:33 & 1706-4346 & 17:06:31.5 & -01:43:46 & 1842-0827 & 18:42:16.1 & -01:08:27\\
0814-2932 & 08:14:02.1 & -01:29:32 & 1213-6233 & 12:13:49.8 & -01:62:33 & 1706-4546 & 17:06:31.5 & -01:45:46 & 1842-1027 & 18:42:16.1 & -01:10:27\\
0814-3132 & 08:14:02.1 & -01:31:32 & 1213-6433 & 12:13:49.8 & -01:64:33 & 1717-3244 & 17:17:59.7 & -01:32:44 & 1850-0030 & 18:50:24.2 & -01:00:30\\
0814-3332 & 08:14:02.1 & -01:33:32 & 1232-6052 & 12:32:40.2 & -01:60:52 & 1717-3444 & 17:17:59.7 & -01:34:44 & 1850-0230 & 18:50:24.2 & -01:02:30\\
0814-3532 & 08:14:02.1 & -01:35:32 & 1232-6252 & 12:32:40.2 & -01:62:52 & 1717-3644 & 17:17:59.7 & -01:36:44 & 1850-0430 & 18:50:24.2 & -01:04:30\\
0814-3732 & 08:14:02.1 & -01:37:32 & 1232-6452 & 12:32:40.2 & -01:64:52 & 1717-3844 & 17:17:59.7 & -01:38:44 & 1850-0630 & 18:50:24.2 & -01:06:30\\
0814-3932 & 08:14:02.1 & -01:39:32 & 1251-6052 & 12:51:30.5 & -01:60:52 & 1717-4044 & 17:17:59.7 & -01:40:44 & 1858-0034 & 18:58:27.3 & -01:00:34\\
0824-3245 & 08:24:55.8 & -01:32:45 & 1251-6252 & 12:51:30.5 & -01:62:52 & 1717-4244 & 17:17:59.7 & -01:42:44 & 1858-0234 & 18:58:27.3 & -01:02:34\\
			\hline
		\end{tabular}
		\normalsize
	\caption{GDS fields and central coordinates. [1] Right ascension, [2] declination, J2000.}
	\label{tab:fields}
\end{table*}

\end{document}